# CCD photometry of distant open clusters:
# II – NGC 6791 [1]

J. Kaluzny

Warsaw University Observatory

Al. Ujazdowskie 4, 00-478 Warsaw, Poland

jka@sirius.astrouw.edu.pl

and

S.M. Rucinski

81 Longbow Square, Scarborough, Ontario M1W 2W6, Canada

affiliated with Department of Astronomy, University of Toronto

and Department of Physics and Astronomy, York University, Toronto, Canada

rucinski@astro.utoronto.ca

## Abstract

Three new photometric CCD-based datasets are presented for NGC 6791. They consist of deep $UBV$ photometry (to $V_{lim} = 24$, $B_{lim} = 24$, $U_{lim} = 23$) of the central parts of the cluster and of selected fields around it, and of relatively shallower UBVI photometry for the whole cluster ($23' \times 23'$). The data have been used to discuss the reddening, metallicity and age of NGC 6791, a cluster which is particularly important because of (1) its most-advanced age among open clusters, (2) metallicity higher than solar and (3) particularly large number of member stars.

We discovered two new very blue stars increasing the total of sdB objects in the cluster to ten. They are located in a very narrow range magnitudes $V = 17.7 \pm 0.5$; their blue colors strongly contrain our determination of the reddening of the cluster to $E(B-V) = 0.17 \pm 0.01$. We detected no other blue stars except a population of faint objects with $V \approx 22$ in the center of the cluster.

Lack of standard data for clusters with metallicity much higher than solar considerably limited the use of the two-color $U - B$ versus $B - V$ diagram to determine metallicity of the cluster. We have used the new theoretical isochrones in a differential determination of metallicity and age of NGC 6791 using M67 and NGC 188 as two clusters with well-known abundaces and age. We find $[Fe/H] \simeq +0.5$ and age by 1 Gyr older than that of NGC 188.

The luminosity function of the cluster is surpisingly flat while the "secondary" main-sequence of equal-mass binaries is weaker than in less-populous old open clusters. We explain both features by the large total mass of the cluster which has lead to weak evaporation of low-mass and single stars from the cluster.

## 1. Introduction

As a candidate for being the oldest open cluster known, NGC 6791 has attracted much interest of observers during the last three decades. Its pioneering study, based on the photographic photometry, was conducted by Kinman (1965). Anthony-Twarog and Twarog (1985) obtained video camera photometry for a relatively small field located near the center of NGC 6791. Janes (1988) and Kaluzny (1990) published CCD BV photometry for fields covering more or less a quarter of the central part of the cluster. Kaluzny & Udalski (1992; hereafter KU) obtained CCD $BVI$ data for most of the area of NGC 6791 as well as for four "comparison" fields located around the cluster, to study global characteristics of NGC 6791. This study resulted also in a discovery of

---

eight very blue stars, candidates for being hot subdwarfs belonging to the cluster. Subsequently, Liebert et al. (1994) obtained spectroscopy for seven brightest candidates among the sdB/O stars. All but one star turned out to be hot subdwarfs and likely members of the cluster. So far, no hot subdwarfs are know in open clusters other than NGC 6791 (except one candidate in NGC 188, cf. Sandage 1962). The uniqueness of NGC 6791 in this respect is most probably due to its relatively high metallicity and/or old age. Recent estimates of the age of NGC 6791 range from $6.5^{+1.5}_{-0.5}$ Gyr (Demarque et al. 1992 ) to 9 Gyr (Garnavich et al. 1994). Meynet et al. (1993) obtained the age of 8.9 Gyr on the basis of the new set of Geneva isochrones. While early photometric studies implied a metallicity of NGC 6791 close to the solar (Harris & Canterna 1981; Geisler et al. 1991), the latest studies indicate that it is in fact slightly above solar ($[Fe/H] = +0.19 \pm 0.19$, Friel & Janes 1993; $[Fe/H] = +0.3$, Garnavich et al. 1994).

Recently, Montgomery et al. (1994) discussed reddening and metallicity of NGC 6791. While this study came after most results of this study were already obtained, we will comment on some of their conclusions at places of an obvious disagreement.

In this paper, we present deep $UBV$ photometry of stars from the central part of NGC 6791 and of stars in a nearby comparison field. These data are used to constrain the cluster parameters and to determine its luminosity function. We obtained also relatively shallower $UBVI$ data for stars within a field of $\approx 23 \times 23$ arcmin$^2$ centered on the cluster center. These data extend and supersede the study conducted by KU in respect of the areal coverage and limiting magnitude. We searched also successfully for new hot-subdwarf candidates in NGC 6791.

## 2. Observations and Reductions

Our data set consist of a large number of CCD frames collected during three observing runs at the Kitt Peak National Observatory. These new data are supplemented by photometry reported by Kaluzny and Rucinski (1993) which we use here to a limited extent.

### 2.1. October 1991 2.1-m + T1KA data

On the nights of 1991 October 5 & 6 (UT), we observed the cluster using the 2.1-m telescope, equipped with a Tektronix 1024×1024 CCD (T1KA) camera. The field of view was about 5.1×5.1 arcmin$^2$ with the scale 0.30 arcsec/pixel. Four partly overlapping fields were observed with the $UBV$ filters. Field F1 covered central part of the cluster. Fields F2 and F3 were located 4′ and 8′ North relatively to field F1. Field F4 was located about 12′ North of the cluster center. Very few − if any − cluster stars are expected to be present in this field (see KU); in a subsequent discussion we will call it "a comparison field". The "central" and "comparison" fields were observed in a very similar way, with the same exposure times, same air-masses and seeing conditions. Several long exposures of fields F1 and F4 were obtained with seeing ranging from 0.9 arcsec to 1.0 arcsec. Only exposures with seeing better than 1.25 arcsec were used for deriving cluster photometry for fields F1 and F4. The exposure times for F1 and F4 were 60 and 600 seconds in $V$, 100 and 900 seconds in $B$ and 1200 seconds in $U$. For fields F2 and F3, we took only long exposures and used them to check the zero points of our photometry.

14 and 15 standard stars from the Landolt (1983) list were observed on the nights of October 5 and October 6, respectively. In addition, we observed twice 7 standards from the field around Ru 149 (Landolt 1992) and several stars from the so called "north consortium field" located near M92 (Stetson & Harris 1988). The transformation equations adopted for the night of October 6 are:

$$v = 2.039 + V + 0.014 \times (B-V) + 0.193 \times (X - 1.25)$$
$$b - v = 0.442 + 0.874 \times (B-V)$$

(1)

$$+0.098 \times (X - 1.25) \qquad (2)$$
$$u - b = 2.246 + 1.167 \times (U - B) - 0.079 \times (U - B)^2$$
$$+0.301 \times (X - 1.25) - 0.03 \times (X - 1.25) \times (U - B) \qquad (3)$$

where $X$ is the airmass and the lower case letters refer to the instrumental magnitudes normalized to 1 sec exposures. The instrumental photometry was extracted with the DAOPHOT/ALLSTAR V2.0 (Stetson 1987, 1991) package. Aperture photometry of standard stars was performed with an aperture radius of 15 pixels (4.5"). For stars in the cluster field, we used profile-fitting photometry. It was found that the point spread function (PSF) showed a small but clearly real position dependence. Application of a linearly variable PSF resolved this difficulty. The actual procedure used to obtain final photometry on the standard system was described in details by Kaluzny (1994). Here we would like to mention only that care was taken to flag stars with relatively poor and/or unreliable photometry. Also, we flagged objects which seemed to be potentially variable.

In Fig. 1, we show residuals between the standard and calculated magnitudes and colors for the standard stars observed on the night of October 6. These standards were observed at air masses spanning the range from 1.19 to 1.93. One can see that the adopted linear transformation reproduces very well the $V$ magnitudes and $B - V$ colors as for most standards the residuals $\delta V$ and $\delta(B - V)$ do not exceed 0.015 mag. The $U - B$ transformation leads to small residuals only for stars with $U - B > 0.7$ and relatively larger residuals are observed for stars with $-0.3 < U - B < 0.2$. We suspect that the problem with the $U - B$ transformation is due to a rapid decline of quantum efficiency of the T1KA chip within the range of wavelengths defined by the $U$ band. The $U - B$ transformation listed above is based only on standard stars with $U - B > -0.3$. The two very blue standards with $U - B \approx -0.9$ exhibit large negative residuals of a similar size (derived colors are too red). This indicates that for very blue stars, the $U - B$ colors derived with the adopted transformation need to be corrected. Accordingly, an offset of $-0.26$ mag was applied to the calculated $U - B$ color of NGC 6791 stars with $U - B < -0.5$.

In Fig. 2 we present color-magnitude diagrams (CMD hereafter) for all stars measured in the observed area of NGC 6791. The external photometric errors are estimated to be 0.015 mag for the $V$ magnitude and $B - V$ color, and 0.03 mag for the $U - V$ color. These estimates are based not only on the overall quality of the transformation (eg. Fig. 1) but also on consistency tests for nights of Oct 5 and Oct 6 which were cross-calibrated with respective transformation coefficients. These latter tests indicate differences in zero points smaller than 0.01 for V and $B - V$ and smaller than 0.02 for U-B.

In Fig. 3 we compare present $BV$ photometry of the central field of NGC 6791 with photometry published by KU. One can see that both sets of photometry are in good agreement and that observed differences are well within ranges implied by the quoted external errors of zero points of photometry. The average differences (present - KU) calculated for stars with $14 < V < 18$ are $\Delta V = -0.018$ and $\Delta(B - V) = +0.009$.

### 2.2. October 1992 0.9-m data

On five nights spanning 1992 October 17 -21 (UT), we observed the cluster using the 0.9-m telescope, equipped with a Tektronix $2048 \times 2048$ CCD (T2KA camera). The field of view was about $23 \times 23$ arcmin$^2$ with the scale of 0.68 arcsec/pixel. Unfortunately, during this run, the 0.9-m telescope produced images with strongly variable PSF. Stellar images were particularly poor for the first 500 columns of the chip. This section of frames was not used for photometry. In addition to the field centered on the cluster, we imaged two fields with centers offset by about 8′ North and about 3′ West, respectively. Some exposures were taken through thin clouds and, for most frames, the seeing was about 2 arcsec. For each field, we took a pair of exposures in V (120 sec and 600 sec), a pair of exposures in $B$ (200 sec and 900 sec) and one exposure in $U$ (1800 sec). One relatively short exposure in $U$ (360 sec) was taken for the field centered on the cluster center. We took also long exposures (600 sec) in $I$ of two of the fields. In Fig. 4 we show a schematic finding chart for the observed field. The instrumental photometry was extracted with the DAOPHOT/ALLSTAR V2.0 (Stetson 1987, 1991) package. The positional variability of the PSF was modeled with a second order polynomial. Despite of these precautions, examination of star-subtracted frames showed presence of a systematic pattern in residual stellar images which implies that our photometry still suffers from some systematic position-dependent errors. The size of these errors is estimated below. Observations of more that 100 standards from Landolt (1992) were used to establish transformation from the instrumental to the standard system. The following relations were adopted:

$$v = 4.001 + V - 0.002 \times (B - V) + 0.25 \times (X - 1.25) \tag{4}$$

$$b - v = 0.966 + 0.966 \times (B - V)$$
$$+ 0.168 \times (X - 1.25) - 0.02 \times (X - 1.25) \times (B - V) \tag{5}$$

$$u - b = 1.796 + 1.010 \times (U - B) + 0.59 \times (X - 1.25)$$
$$- 0.02 \times (X - 1.25) \times (U - B) \tag{6}$$

$$v - i = -0.721 + 1.026 \times (V - I) + 0.068 \times (X - 1.25) \tag{7}$$

In Fig. 5 we show the color-magnitude diagrams (CMD) $V$ vs. $B - V$ and $V$ vs. $U - B$ for the whole observed field. $B - V$ color was derived for 11964 objects with $V < 21.0$ while $U - B$ color was derived for 6737 objects with $V < 19.5$. In Fig. 6 we show $V$ vs. $V - I$ CMD. Almost 13 thousand stars are plotted in this figure. Because of the lack of short exposures for the $I$ band, our $VI$ photometry does not include any objects brighter than $I \approx 13$. In particular we did not obtain $VI$ photometry for the very red giants belonging to NGC 6791 which were discussed recently by Garnavich et al. (1994). However, we derived $UBV$ photometry for most of these giants and the $BVI$ data for several of them had been published (contrary to the claim made by Garnavich et al. (1994)) by KU. As we mentioned above, the UBVI photometry obtained during this run suffers from some systematic position dependent errors. The size of these errors can be appreciated in Fig. 7 which shows a comparison of the present data with that published by KU.

### 2.3. October 1991 2.1-m ST1K data

On three nights spanning the period 1991 October 2-4 (UT), we observed the cluster using the 2.1-m telescope, equipped with a ST1K $1024 \times 1024$ CCD camera. The field of view was about $4.3 \times 4.3$ arcmin$^2$ with the scale 0.26 arcsec/pixel. Four partly overlapping fields were observed with the $UBV$ filters. This mosaic was centered on the cluster center and covered a field $8.0 \times 8.3$ arcmin$^2$. At least two exposures in each of three filters were obtained for each field. The exposure time was set to 600 sec for all frames. The seeing was good and ranged from 0.95 to 1.3. The data were reduced using procedure analogous to that described in Sec. 2.1. During our run the ST1K camera suffered from a charge transfer problem. The problem lead to formation of "shadow-like" traces on one side of bright stars present in a frame, and affected photometry of faint stars located in vicinity of bright objects. The instrumental photometry was transformed to the standard $UBV$ system using transformation determined based on observations of several standard fields from Landolt (1992). We show the derived CMD's in Fig. 8. One can see that photometry obtained with the 2.1-m telescope and the ST1K chip is comparable in its depth to photometry obtained with the 0.9-m telescope (see Sec. 2.2).

## 3. Hot subdwarfs and cataclysmic variables

As we mentioned already in the Introduction, NGC 6791 harbors several hot stars. Eight candidates for hot subdwarfs were identified in the cluster field by KU. Subsequently Kaluzny and Rucinski (1993) discovered variability of one of these objects, namely the star identified as B7. Their suggestion about the binary nature of B7 was confirmed by Liebert et al. (1994) who obtained spectra of candidates B1-B7. The brightest and coolest candidate, B1, turned out to be, most likely, a non-member B star, while stars $B2 - B6$ are hot subdwarfs and likely members of the cluster. Presence of several sdB/O stars in NGC 6791 indicates that most hot subdwarfs belonging to the galactic disk are formed by its old metal-rich component.

Table 1: UBVI photometry of faint blue stars located in and around the area of NGC 6791. This photometry is based on the data collected in October 92 with the 0.9-m telescope. The quoted formal mean errors (sigma) do not include some possible systematic position-dependent errors of the photometry.

| ID | V | $\sigma V$ | B-V | $\sigma B - V$ | U-B | $\sigma U - B$ | V-I | $\sigma V - I$ |
|----|----|----|----|----|----|----|----|----|
| B1 | 16.918 | 0.006 | 0.100 | 0.010 | -0.079 | 0.013 | 0.082 | 0.011 |
| B11 | 21.322 | 0.077 | 0.424 | 0.089 | -0.434 | 0.084 | | |
| B10 | 16.094 | 0.005 | 0.178 | 0.009 | -0.434 | 0.010 | 0.405 | 0.010 |
| B12 | 21.585 | 0.068 | 0.124 | 0.079 | -0.380 | 0.068 | | |
| B13 | 20.996 | 0.050 | 0.679 | 0.077 | -1.154 | 0.150 | 0.447 | 0.150 |
| B3 | 17.752 | 0.005 | -0.100 | 0.009 | -0.567 | 0.009 | -0.122 | 0.010 |
| B6 | 17.934 | 0.006 | -0.084 | 0.008 | -0.614 | 0.010 | -0.113 | 0.013 |
| B7 | 18.136 | 0.007 | 0.198 | 0.009 | -0.338 | 0.008 | | |
| B14 | 21.380 | 0.092 | 0.795 | 0.114 | -0.496 | 0.121 | 1.072 | 0.109 |
| B15 | 20.271 | 0.017 | 0.104 | 0.022 | -0.470 | 0.026 | 0.159 | 0.067 |
| B5 | 17.880 | 0.005 | -0.096 | 0.008 | -0.559 | 0.008 | -0.118 | 0.011 |
| B16 | 20.564 | 0.028 | 0.571 | 0.036 | -0.371 | 0.044 | 0.803 | 0.057 |
| B4 | 17.869 | 0.005 | -0.131 | 0.008 | -0.600 | 0.008 | -0.153 | 0.014 |
| B17 | 21.520 | 0.060 | -0.352 | 0.067 | -0.789 | 0.035 | | |
| B8 | 20.223 | 0.023 | -0.225 | 0.028 | -0.631 | 0.024 | 0.776 | 0.044 |
| B18 | 21.781 | 0.078 | 0.137 | 0.093 | -0.508 | 0.081 | | |
| B9 | 18.176 | 0.009 | -0.153 | 0.013 | -0.674 | 0.013 | -0.168 | 0.028 |
| B2 | 17.428 | 0.006 | -0.118 | 0.008 | -0.728 | 0.010 | -0.101 | 0.013 |
| B19 | 19.005 | 0.020 | 0.775 | 0.030 | -0.366 | 0.081 | 0.785 | 0.033 |

The photometric data obtained with the 0.9-m telescope in October 92 supersede BVI photometry published by KU in respect of the covered area as well as of the limiting magnitude. Moreover, the new data include the $U - B$ color, what should allow less ambiguous identification of hot stars present in the cluster field. Table 1 contains $UBVI$ photometry for blue stars selected from the data shown in Fig. 5. For $V < 20$, the stars selected had $U - B < -0.2$ (the only exception here is the star B1 with $U - B = -0.079$) while for $V \geq 20$, the selection threshold was shifted to $U - B < -0.4$. In the case of stars with $V \geq 20$, an additional condition was that the star had to be detected on at least two B frames and at least two U frames. It is worth to note that the distribution of colors for the field stars shows a sharp cut-off at $U - B \approx 0.1$ and $B - V \approx 0.6$ (see Fig. 5) corresponding to the location of the turnoff of the field stars from the disk and halo; therefore, we feel that our selection process was free of any ambiguity. Table 1 includes all eight candidates identified by KU. Star B1 has much redder color $U - B$ than stars B2-B8. This supports conclusion of Liebert et al. (1994) that it is not a hot subdwarf.

Two new candidates for hot subdwarf were identified among stars with $V < 20.0$. Star B9 joins a clump of six previously identified sdB/O stars with $V \approx 17.8$ and $B - V \approx -0.1$. This star was located outside field surveyed by KU. Star B10 was observed by KU but was not selected as a candidate for being a hot subdwarf due to its relatively red $B - V$ color. Examination of data obtained with the 2.1-m telescope shows that B10 is in fact a blend of two stars with $\Delta V \approx 2.0$ mag and angular separation $d \approx 0.7$ arcsec. Both components were measured together on the frames collected with the 0.9-m telescope. As we discuss bellow the brighter component of this close pair is a candidate for being a hot subdwarf.

Table 2: Rectangular and equatorial coordinates of blue stars identified in the field in and around NGC 6791. Rectangular coordinates can be used to identify all objects on the image included in the set of supplementary data accompanying this paper (see Appendix A).

| No | X | Y | RA(2000) | DEC(2000) |
|---|---|---|---|---|
| B1  | 386.86  | 1306.20 | 19:20:40.33 | 37:53:50.9 |
| B11 | 624.83  | 1472.04 | 19:20:30.78 | 37:51:06.2 |
| B10 | 657.32  | 936.77  | 19:21:01.92 | 37:50:46.2 |
| B12 | 738.66  | 1876.60 | 19:20:07.31 | 37:49:45.7 |
| B13 | 746.60  | 224.66  | 19:21:43.34 | 37:49:46.9 |
| B3  | 764.04  | 1225.24 | 19:20:45.19 | 37:49:31.5 |
| B6  | 868.57  | 1223.24 | 19:20:45.34 | 37:48:19.5 |
| B7  | 904.13  | 843.76  | 19:21:07.41 | 37:47:56.5 |
| B14 | 963.46  | 987.43  | 19:20:59.08 | 37:47:15.1 |
| B15 | 984.80  | 1644.74 | 19:20:20.90 | 37:46:57.4 |
| B5  | 985.98  | 913.86  | 19:21:03.36 | 37:46:59.8 |
| B16 | 1012.59 | 1231.55 | 19:20:44.92 | 37:46:40.2 |
| B4  | 1086.27 | 750.16  | 19:21:12.91 | 37:45:51.3 |
| B17 | 1112.93 | 1891.98 | 19:20:06.60 | 37:45:27.8 |
| B8  | 1168.30 | 1390.61 | 19:20:35.74 | 37:44:52.3 |
| B18 | 1174.78 | 1714.76 | 19:20:16.92 | 37:44:46.2 |
| B9  | 1246.52 | 197.87  | 19:21:45.03 | 37:44:02.4 |
| B2  | 1450.29 | 1148.33 | 19:20:49.92 | 37:41:39.0 |
| B19 | 1992.16 | 927.89  | 19:21:02.93 | 37:35:26.4 |

Table 3: Average magnitude and colors for star B8 observed during four observing runs at the Kitt Peak National Observatory.

| Telescope | UT Dates | $<V>$ | $<B-V>$ | $<U-B>$ |
|-----------|----------|-------|---------|---------|
| 0.9-m | Jun 5, 1991 | 19.91 | +0.14 | |
| 2.1-m | Oct 2-4, 1991 | 20.68 | -0.10 | -.68 |
| 0.9-m | Oct 8-11, 1991 | 20.66 | +0.21 | |
| 0.9-m | Oct 1992, | 20.22 | -0.22 | -0.63 |

In Table 2 we list rectangular and equatorial coordinates of all stars from Table 1. An astrometric solution was based on position of 55 stars from the GSC (Lasker et al. 1988). This solution reproduces coordinates of GSC stars with residuals not exceeding 1 arcsec. However, comparison of equatorial coordinates of B1 − B8 in Table 2 with coordinates published by KU shows differences reaching 4 arcsec. The problem is caused by the fact that all GSC stars used for astrometric solution were located outside central part of the cluster. Distortion of images (see Sec. 2.2) obtained with the 0.9-m telescope leads to relatively large errors of equatorial coordinates derived for stars located close to the central part of the observed field. In Fig. 9 we present finding charts allowing identification of stars B9 and B10 (blue object is a brighter component of a close visual pair). Charts for stars B1 − B8 were published by KU. The remaining blue objects (B11 − B19) can be identified on an image included in supplementary data submitted with this paper (see Appendix A). Examination of colors of the star B8 listed in Table 1 suggest that this star – like B7 – is a binary system. In Table 3 we list average V magnitude and average colors of B8 measured during four observing runs at KPNO. It is evident that B8 is a variable star. In particular, during four nights spanning the period of October 2-4, 1991 (Kaluzny and Rucinski 1993), it exhibited a range of magnitudes from $V = 20.31$ to $V = 20.82$. Variability of B8 and its very blue $U - B$ color indicate that it is most probably a cataclysmic variable. The main argument in favor of the cluster membership of B8 is its location in the central part of the cluster. For the assumed cluster membership, the absolute magnitude of B8 would be $M_v \approx 7$. Such an absolute magnitude is typical for U Gem-type cataclysmic variables as well as for polars (AM Her stars). So far, the only known cataclysmic variable being a highly probable member of any open cluster was located in M67 by Gilliland et al. (1991).

It is remarkable that hot subdwarfs in NGC 6791 show such a small range of magnitudes and colors. It should be stressed that no blue objects are present in the CMD shown in Fig. 5 within the range entirely accessible to our photometry $18.2 < V < 20.2$. This deficiency is certainly not due to any selection effect as several fainter blue objects with $V > 20.2$ ($M_v > 6.7$) were detected (eg. Table 2). It is difficult however to assess how many of these faint blue stars are also cluster members. We may note only that Drukier et al. (1989) discovered a sequence of faint hot subdwarfs in the metal-rich disk globular cluster M71. A significant fraction of objects with $V > 20$ in Table 2 may be actually quasars. The surface density of quasars with $B < 22$ is about 70 per deg$^2$ (Boyle at al. 1991). Our survey covered about (1/9) deg$^2$ so we may expect detection of less than about 7 quasars with $V < 21$ and $B - V < 0.5$. It would be useful to obtain information about variability of the faint blue objects in the field of NGC 6791 as this might, potentially, resolve the question of their nature through comparison with variability of quasars

Table 4: UBVI photometry of faint blue stars located in and around the area of NGC 6791. This photometry is based on the data collected on October 91 with the 2.1-m telescope and described in Sec. 2.1. The formal errors quoted for the $U - B$ colors do not reflect uncertainty involed in the UB transformation for stars with $U - B < -0.5$ (see Sec. 2.1 and Fig. 1)

| ID | V | $\sigma V$ | B-V | $\sigma B - V$ | U-B | $\sigma U - B$ |
|-----|--------|-------|--------|-------|--------|-------|
| B20 | 22.113 | 0.021 | 0.045  | 0.033 | -0.761 | 0.045 |
| B21 | 22.002 | 0.023 | 0.600  | 0.038 | -0.114 | 0.093 |
| B22 | 21.813 | 0.040 | -0.116 | 0.050 | -0.934 | 0.027 |
| B23 | 23.217 | 0.074 | -0.025 | 0.100 | -1.237 | 0.275 |
| B24 | 20.551 | 0.007 | 0.596  | 0.014 | -0.334 | 0.020 |
| B5  | 17.904 | 0.002 | -0.105 | 0.005 | -0.811 | 0.004 |
| B7  | 18.055 | 0.003 | 0.145  | 0.007 | -0.364 | 0.005 |
| B6  | 17.970 | 0.002 | -0.102 | 0.005 | -0.861 | 0.004 |
| B3  | 17.775 | 0.002 | -0.117 | 0.005 | -0.816 | 0.003 |
| B10 | 16.278 | 0.010 | 0.014  | 0.014 | -0.431 | 0.006 |
| B1  | 16.968 | 0.005 | 0.070  | 0.010 | -0.177 | 0.009 |
| B26 | 21.313 | 0.013 | 0.350  | 0.027 | -0.847 | 0.032 |
| B27 | 21.776 | 0.024 | 0.383  | 0.045 | -0.471 | 0.071 |

taking place typically in time-scales of months.

Now we turn to the data collected with the 2.1-m telescope which are described in Sec. 2.1. In Table 4 we give $UBV$ photometry for all stars in Fig. 2 with $U - B < -0.1$. All these objects can be identified unambiguously using data described in Appendix A. As we discussed above, the external errors of the new $BV$ photometry are unlikely to be greater than 0.015 mag. The external errors of U-B colors of extremely blue stars are likely to be about 0.05 mag. It is worth to note a very small spread in $B - V$ exhibited by stars B3, B5 and B6.

In Fig. 10 the CMD's $V$ vs. $B - V$ for fields F1 (cluster center) and F4 (comparison field). The areas covered by both fields were equal to and the limiting magnitude of the photometry were very similar. One can note an apparent excess of faint blue stars with $V > 22$ in the CMD of the field F1 in respect to CMD of the field F4. This suggests that at least a fraction of faint blue stars observed in the cluster center are members of NGC 6791. These stars seem to be too numerous to be bright white dwarfs and might be descendents of bright sdB/O stars. On the other hand, some of them may be simply extragalactic objects.

## 4. Cluster Parameters

### 4.1. Reddening and metallicity

During last few years several authors attempted to determine age and other parameters of NGC 6791 (Carraro et al. 1994; Demarque et al. 1992; Garnavich et al. 1994; Meynet et al. 1993). The main obstacle affecting precise determination of cluster parameters was uncertainty concerning its reddening. Previous estimates of $E(B-V)$ based on multicolor photometry ranged from 0.10 (Janes 1984) to 0.22 (Kinman 1965). Comparison of old photoelectric photometry with more recent CCD photometry indicates that the former suffered from large random and/or systematic errors (e.g. KU). Recently Montgomery et al. (1994) combined their new $UBV$ photoelectric photometry of 17 stars from the NGC 6791 area with photometry of 14 stars published by Harris & Canterna (1981) and derived for this sample $E(B - V) = 0.10 \pm 0.02$.

Table 5: Values $\delta(U-B)$ derived for different assumed values of $E(B-V)$. The ultraviolet excess was derived separately for giants and for two groups of stars from the upper main sequence. See text for more details.

| E(B-V) | giants | $17.5 < V < 18.9$ | $18.9 < V < 19.4$ |
|--------|--------|-------------------|-------------------|
| 0.05 | -0.021 | -0.003 | +0.017 |
| 0.08 | -0.050 | -0.029 | -0.019 |
| 0.11 | -0.078 | -0.056 | -0.052 |
| 0.13 | -0.098 | -0.077 | -0.072 |
| 0.15 | -0.118 | -0.098 | -0.094 |
| 0.17 | -0.138 | -0.119 | -0.113 |
| 0.19 | -0.158 | -0.134 | -0.132 |
| 0.21 | -0.179 | -0.166 | -0.148 |

Our data permit to estimate reddening of NGC 6791 in two ways. The first approach exploits the fact that the cluster contains several hot subdwarfs. KU already noted that colors of these subdwarfs set an upper limit on the reddening: $E(B-V) < 0.195$. Later on Liebert et al. (1994) determined unreddened colors of four sdB stars belonging to NGC 6791 and derived $E(B-V) = 0.143 \pm 0.021$. In this study, we obtained new precise photometry for 3 sdB stars from the sample analyzed by Liebert et al. (1994). As we discussed in Sec.2.1., photometry obtained with the 2.1-m telescope and T1KA camera leads to $B-V$ colors bluer by 0.009 in respect to the KU data. Using photometry of B3, B5 and B6 listed in Table 4 and unreddened colors given by Liebert et al. (1994), we derive $< E(B-V) >= 0.168 \pm 0.013$.

Our second estimate of reddening of NGC 6791 is based on a more traditional use of our $UBV$ photometry described in Sec. 2.1. In principle, having in hand a precise $UBV$ photometry for main sequence stars and giants, one may derive simultaneously cluster reddening and its metallicity (eg. Cameron 1985). In practice, for very old clusters, it is difficult to separate the colors shifts due to interstellar reddening from those due to non-solar metallicity. In the present analysis we calculated $\delta(U-B)$ for three groups of stars from NGC 6791:

- cluster giants

- stars from turnoff region of the CMD with $17.5 < V < 18.9$

- stars from main sequence with $18.9 < V < 19.4$

Each value of $\delta(U-B)$ was calculated for several assumed values of $E(B-V)$. The detailed description of the procedure can be found in Kaluzny and Mazur (1991a,b). The results are given in Table 5.

In Fig. 11 we show the color-color diagram for the three groups of stars used for calculation of $\delta(U-B)$ together with the standard relation for Hyades (Sandage & Eggen 1959; Eggen 1966). The standard relation was corrected for reddening of $E(B-V) = 0.17$. Examination of Table 5 shows that values of $\delta(U-B)$ derived for giants are only marginally lower than values of $\delta(U-B)$ derived for main sequence stars. In fact, evolved stars from cluster turnoff are expected to show slightly lower ultraviolet excesses than main sequence stars located below the turnoff point. Negative values of $\delta(U-B)$ observed for $E(B-V) \geq 0.08$ imply metallicity higher than solar. For $E(B-V) = 0.17$ (a value obtained for sdB stars in NGC 6791), we obtained $\delta(U-B) \approx -0.13$. Unfortunately none of the available calibrations between $[Fe/H]$ and $\delta(U-B)$ apply to stars with metallicity significantly higher than solar metallicity. Extrapolation of the relation given by Cameron (1985) results in $[Fe/H] = +0.49$ for $\delta(U - B) \approx -0.13$, with $[Fe/H] = +0.08$ implied for the Hyades. If, following Friel & Merchant Boesgard (1992), we adopt $[Fe/H] = +0.127$ for Hyades then metallicity of NGC 6791 is increased to $[Fe/H] = +0.54$. Friel & Janes (1993) have obtained recently $[Fe/H] = +0.19 \pm 0.19$ from the spectroscopy of several members of NGC 6791. Garnavich et al. (1994) used VI photometry to demonstrate that morphology of the red giant branch of NGC 6791 implies a metallicity much larger than the solar. These authors measured also the equivalent widths of the Ca II lines in the spectra of several cluster giants. Based on these measurements they estimated metallicity of NGC 6791 at $[Fe/H] \approx +0.22$.

As one more evidence against solar metallicity of NGC 6791 we show in Fig. 12 the fiducial sequence of NGC 6791 superimposed on the CMD of M67 (Montgomery et al. 1993). The CMD of M67 has been shifted vertically by $\delta V = 3.93$ and horizontally by $\delta(B - V) = 0.23$. These shifts were calculated assuming that metallicities of both clusters are equal. The shift of $\delta V = 3.93$ is needed to obtain separation of horizontal branches of both clusters equal to 0.08 mag. Such a separation can be expected on the basis of the isochrones published recently by Bertelli et al. (1994, hereinafter called the "Padova isochrones"), if we assume that ages of M67 and NGC 6791 are 5 Gyr and 7.2 Gyr, respectively (Carraro et al. 1994; section 4.2 of this paper). The shift of $\delta(B - V) = 0.23$ is needed to match the unevolved parts of the main sequences of both clusters. Such a shift implies that NGC 6791 shows a differential reddening of $\Delta E(B - V) = 0.23$ relative to M67. The reddening of M67 is $E(B - V) = 0.05$ (Montgomery et al. 1993) so that the total reddening of NGC 6791 would then be $E(B - V) = 0.28$. Examination of Table 5 show that such a reddening would correspond to an unacceptably large and negative value of $\delta(U - B)$ and, consequently, to a much higher than solar metallicity of the cluster. Reddening as high as $E(B - V) = 0.28$ would be also inconsistent with the observed colors of sdB/O stars belonging to NGC 6791.

It is possible to estimate differential metallicity of NGC 6791 relative to M67 by comparing the CMD's of both clusters. In our analysis we used calibrations which are based on the theoretical isochrones published by Bertelli et al. (1994). These models permit to determine average $M_V$ of the horizontal branch for a given age and metallicity of a cluster. For M67, we adopted $[Fe/H] = -0.08$ (Friel and Janes 1993) and age equal to 5 Gyr (Carraro et al. 1994) which implies $M_V = 0.87$ for the average absolute magnitude of helium burning giants. Let us adopt for the time being $[Fe/H] = +0.30$ for NGC 6791. The age of the cluster is about 7.2 Gyr on the scale of the Padova isochrones (see Sec. 4.2). These parameters imply $M_V = 1.06$ for the average location of helium burning giants in NGC 6791. We adopted also $E(B - V) = 0.17$ for NGC 6791 (this section) and $E(B - V) = 0.05$ for M 67 (Montgomery et al. 1993). In Fig. 13 we show the CMD's for both clusters. The CMD of M67 was shifted by $\delta V = 3.84$ and $\delta(B - V) = 0.17 - 0.05 = 0.12$. The shift of $\delta V = 3.84$ results in separation of horizontal branches by $\delta M_V = M_V(N6791) - M_V(M67) = 1.06 - 0.87 = 0.19$. The unevolved parts of main sequences of both clusters are separated by $\delta V \approx 0.30$ at $(B - V) \approx 1.05$ or $(B - V)_0 \approx 0.88$. Such a separation implies difference of metallicities $\Delta Z = Z_{NGC6791} - Z_{M67} = 0.021$. Resulting metallicity of NGC 6791 is then $Z = 0.038$ or $[Fe/H] = 0.29$ which is self-consistent with assumption made at the beginning of this paragraph. Adoption of a lower value of $[Fe/H]$ requires adoption of a smaller separation between horizontal branches of M67 and NGC 6791 and a larger value of the $\delta V$ shift. This in consequence leads to a larger separation between unevolved main sequences of both clusters. However, larger separation of unevolved main sequences implies in turn a larger difference of metallicities. Hence, only adoption of $[Fe/H] = +0.30$ for NGC 6791 − or more precisely $\Delta[Fe/H] = 0.38$ between NGC 6791 and M67 − leads to self consistent solution. In Fig. 14 we present comparison of CMDs of M67 and NGC 6791 based on adoption of $[Fe/H] = 0.19$ and $E(B - V) = 0.10$ (Montgomery et al 1994) for the latter cluster. Using the Padova isochrones, we found that the predicted separation of horizontal branches is then $\Delta M_V = 0.15$. The resulting separation of unevolved main sequences of both clusters is then $\delta V = 0.72$ implying differential metallicity $\Delta Z = 0.050$ and finally $[Fe/H] = 0.54$ for NGC 6791. Clearly adoption of $E(B - V) = 0.10$ leads to an even larger difference between metallicities of M67 and NGC 6791 than adoption of $E(B - V) = 0.17$.

Recently Montgomery et al. (1994) combined their new $UBV$ photoelectric photometry of 17 stars from the NGC 6791 area with photometry of 14 stars published by Harris & Canterna (1981) and derived for this sample $E(B - V) = 0.10 \pm 0.02$. While looking at the possible source of discrepancy between results of Montgomery et al. (1994) and our determination $E(B - V) = 0.17$ we noted that a significant fraction of stars observed by them are foreground objects, which are likely located at distances of only a few hundred parsecs from the Sun. This conclusion is based on comparison of apparent $V$ magnitudes listed in Table II in Montgomery et al. (1994) with absolute magnitudes derived from relation $M_V = M_V(B - V)$ for the dwarfs of solar metallicity. In our opinion $E(B - V) = 0.10$ is in fact just a firm lower limit on the reddening of NGC 6791.

### 4.2. Age

The issue of the age of NGC 6791 was a subject of vigorous activity in the last few years. Garnavich et al. (1994) obtained an age of about 9 Gyr based on the direct isochrone fitting to the cluster giant branch. Demarque et al. (1992) suggested that the age of NGC 6791 can be comparable to the age of NGC 188 if the former cluster is more metal rich than the latter. In Fig. 15 we show the fiducial sequences for NGC 6791 and NGC 188. The relation for NGC 188 is based on photometry by Kaluzny (1990) supplemented with unpublished photometry derived at the KPNO by the same author. Offsets of $\delta V = 2.21$ and $\delta(B - V) = 0.18$ were applied to the NGC 188 sequence to match the colors of turnoffs of both clusters as well as levels of their horizontal branches. In the case of NGC 188, our data included the two He burning (Red Clump) giants. The V magnitudes of these giants differ by 0.08 mag and the fainter one is located at $V = 12.38$. Their average magnitude $< V_{clump} >\, = 12.34$ agrees very well with $< V_{clump} >\, = 12.35$ which is the value reported by Twarog & Anthony-Twarog (1989).

In the discussion below, we will use a parameter $\Delta V$ which is defined as a difference in magnitudes between the average magnitude of the helium burning giants and the brightest point on the main sequence (Anthony-Twarog and Twarog 1985). This parameter is known to correlate with the cluster age but it is also a function of metallicity. We derived $\Delta V = 2.49$ and $\Delta V = 2.83$ for NGC 188 and NGC 6791, respectively. The metallicity of NGC 188 is known to be close to solar; Hobbs et al. (1990) derived $[Fe/H] = -0.08$ which corresponds to $Z = 0.017$ on a scale with $Z_\odot = 0.020$. For NGC 6791 we adopt $Z = 0.039$ (see section 4.1). In Fig. 16 we show lines of constant metallicity on a plane *age* versus $\Delta V$. The relations for $Z = 0.05$, $Z = 0.039$ and $Z = 0.017$ are based on Padova isochrones (Bertelli et al. 1994) with those for $Z = 0.039$ and $Z = 0.017$ obtained by interpolation. Positions of NGC 188 and NGC 6791 in Fig. 16 imply ages equal to 6.0 Gyr and 7.2 Gyr, respectively. Clearly, the derived absolute ages of both clusters are subject to uncertainties specific to a particular set of used isochrones. However, it seems safe to state that NGC 6791 is by about 1 Gyr older than NGC 188.

### 4.3. Distance modulus

Two methods are commonly used for determination of apparent distance moduli of stellar clusters. The first method is based on a comparison of the observed magnitude of the Red Clump giants with their predicted absolute magnitude. Such absolute magnitudes can be obtained either from theoretical isochrones or from observational calibrations. The second method is based on fitting of the unevolved parts of the main sequence with the empirically determined ZAMS or with the empirically calibrated theoretical ZAMS (eg. VandenBerg & Poll 1989). The first method utilizes an assumption that helium burning giants have the same absolute magnitude $M_V$ for clusters older than about 1 Gyr. This was first noted by Cannon (1970) who obtained photometry for several intermediate-age and old open clusters. Recently Castellani et al. (1992) concluded, on the basis of solar metallicity isochrones, that "for clusters having an age greater than 500 Myr, the lower envelope of the He-burning clump is expected at $M_V = 0.85 \pm 0.1$ mag". Theoretical models indicate that $M_V$ of the helium-burning giants is an increasing function of metallicity (eg. Bertelli et al. 1994). The main source of uncertainty concerning properties of helium-burning giants in open clusters are problems with transforming the model quantities $M_{bol}$ and $T_{eff}$ into the observational plane $M_V$ vs. $B - V$.

It is worth to emphasize, that distance moduli based on the main-sequence fitting are known with an accuracy better than 0.2-0.3 mag for very few intermediateage and old open clusters . For most well-studied clusters the main obstacles in determining precise value of unreddened distance moduli are uncertainties concerning reddening and/or metallicity.

After these cautionary remarks we may turn back to NGC 6791. The average observed magnitude of the helium-burning giants in this cluster is $< V > = 14.57$. According to the Padova isochrones (Bertelli et al. 1994), for clusters of solar composition, the helium-burning giants have $< M_V >$ ranging from 0.88 to 0.97 for ages ranging from 4.0 to 7.9 Gyr, respectively. An increase of metallicity from $Z = 0.02$ to $Z = 0.05$ leads to an increase of $< M_V >$ by about 0.15 mag for a fixed age. Consequently, we may assume that for ages ranging from 4 to 8 Gyr and $Z$ ranging from 0.02 to 0.05, $< M_V >$ should be within 0.88 to 1.12. Making a rather safe assumption that the age and metallicity of NGC 6791 fall into the above ranges, we arrive at a conclusion that an apparent distance modulus of the cluster is confined within an interval $\approx 13.45$ to $\approx 13.7$. In the previous sections we obtained for NGC 6791 $Z = 0.039$ and the age $\approx 7.2$ Gyr. For these values, the Padova isochrones imply $< M_V > = 1.05$. Consequently the apparent distance modulus of the cluster would be $(m - M)_V = 13.52$ mag. Adopting further $E(B - V) = 0.17$, we obtain $(m - M)_0 = 12.97$ which corresponds to a heliocentric distance of $d = 3.9$ kpc. Admittedly, this result does not differ substantially from several earlier estimates (Kaluzny 1989; KU; Garnavich et al. 1994; Carraro et al. 1994; Demarque et al. 1992). The only exception is the recent results of Montgomery et al. (1994) who obtained $(m - M)_V = 12.96$ and $(m - M)_0 = 12.66$. Their estimate is clearly incorrect. The error resulted from confusing the difference of $M_V$ for stars of the same mass but different metallicities with the difference of $M_V$ measured at a fixed color between isochrones for different metallicities.

Our photometry of NGC 6791 described in Sec. 2.1 extends well into the unevolved part of the cluster main-sequence. Thus, it could be used for determination of the distance modulus through the ZAMS fitting, after adoption of the values for $E(B - V)$ and $Z$ that we determined earlier. This way we could, in principle, observationally determine the value of $< M_V >$ for the helium-burning giants in this metal-rich cluster. Such a procedure would in fact involve a circular argument as we assumed a specific relation between $< M_V >$ and metallicity for the Red Clump giants to determine the cluster metallicity (see Sec. 4.2). However, it is permissible to simply use ZAMS fitting method to determine distance moduli of the cluster for some *assumed* values of its metallicity.

By interpolation in the Padova isochrones for $Z = 0.02$ and $Z = 0.05$, we obtained the location of ZAMS for $Z = 0.039$ (corresponding to $[Fe/H] = +0.30$), the value derived in Sec. 4.1., and for $Z = 0.031$ ($[Fe/H] = +0.20$, the value obtained spectroscopically by Friel & Janes (1993) and by Garnavich et al. (1994)). Using the observed location of unevolved NGC 6791 stars with $B - V \approx 1.15$, we obtained $(m - M)_V = 13.48$ and $(m - M)_V = 13.37$ for $Z = 0.039$ and $Z = 0.031$, respectively. These estimates were obtained for the adopted $E(B - V) = 0.17$. Adoption of a lower reddening, as advocated by Montgomery et al. (1994), leads to larger values of apparent distance moduli. We note paranthetically, that the solar metallicity ZAMS, based on Padova isochrones, is brighter by about 0.15 mag for $0.7 < B - V < 1.0$ than the empirically calibrated ZAMS of VandenBerg and Poll (1989).

Application of the procedure described by VandenBerg and Poll (1989) leads to $(m - M)_V = 13.30$ for $[Fe/H] = +0.20$, and $(m - M)_V = 13.41$ for $[Fe/H] = +0.30$. The helium content for a given $Z$ was calculated assuming the relation $\Delta Y/\Delta Z = 2.5$ and $Y_\odot = 0.27$.

It may be concluded that all of the above estimates of the distance modulus of NGC 6791 are close to each other. For $[Fe/H] = +0.20$, we obtained $13.30 < (m - M)_V < 13.37$ while for $[Fe/H] = +0.30$, we obtained $13.41 < (m - M)_V < 13.52$.

### 4.4. Luminosity function

It is known that several old open clusters show a dramatic deficiency of lower main sequence stars ($M_V > 7$) relative to the luminosity function of the field stars. This is interpreted as a result of a selective "evaporation" of low-mass stars from the host cluster. On the other hand, luminosity functions of globular clusters tend to rise toward very faint magnitudes (e.g. Fahlman et al. 1989). NGC 6791 is a very massive object in comparison with other well studied old open clusters (KU; Kinman 1965). It is therefore interesting to check whether its luminosity function is typical for open or globular clusters. Our analysis is based on photometry described in Section 2.1. We limited our attention to fields F1 (cluster center) and F4 (comparison field) since photometry of comparable quality and deepness was obtained for these two fields. Instead of using combined photometry, we selected the best available frame for each filter/field combination. An excellent discussion of problems associated with determination of luminosity function can be found in Stetson and Harris (1988) and Bergbusch (1993). The rather low crowding of the analyzed fields justifies the relatively simple approach which we describe below.

In Fig. 17 we present the observed luminosity functions for fields F1 and F4. The simplified relation in the form $V = v_{inst} + const$ was applied to transform instrumental magnitudes to the standard system (analogous relations were used for the $B$ and $U$ filter frames). Systematic errors due to neglecting of color terms in adopted transformations do not exceed 0.03 mag in the case of the $V$ magnitudes and 0.15 mag in the case of the $B$ and $U$ magnitudes (see Eqs. 1-3). Several selection criteria were applied while deriving the luminosity function. These criteria make use of three parameters returned by DoPHOT for each object with measured magnitude. The parameter $SIGMA$ is a formal error of the profile magnitude and the parameter $CHI1$ measures the quality of the model PSF fit to the object's profile. The parameter $CHI2$ describes the shape of the measured object and allows to distinguish between stellar-like object and extended objects (like galaxies) or abnormally compact objects (like cosmic-ray traces). As an example, we show in Fig. 18 dependence of all three parameters on the V magnitude for the 600 sec image of field F4. Only objects with $CHI1 < 2$ and $-1 < CHI2 < 1$ were counted while deriving the luminosity function. In addition, objects deviating strongly from the average relation $SIGMA(mag)$ were rejected. All objects located closer than 15 pixels from the edge of a frame were also excluded.

To correct the derived LF for the incompleteness of photometry, we performed a test based on addition of artificial stars. Ten images, each containing 50 artificial stars added to the original frames, were created for each 0.5 magnitude bin. This gave results based on statistics for 500 added stars, which was adequate for estimation of incompletness corrections. The same criteria as those applied to the original images were used while counting numbers of the recovered artificial stars. In Figs. 19 and 20, we show luminosity functions corrected for the incompleteness of the photometry. The observed and rectified luminosity functions are given in Tables 6 and 7. In Fig. 21 we show a differential luminosity function in V. It was obtained by subtraction of luminosity function for field F4 from the luminosity function of field F1 (Figs. 19 and 20). It is visible that luminosity function of field F1, corrected for incompleteness of photometry and background/foreground contamination, does not show any steep decline at its faint end. For the adopted distance modulus $(m - M)_V = 13.5$ the last bin of presented V luminosity function

Table 6: Luminosity function for the central field of NGC 6791 (field F1). The first two columns define ranges of magnitudes for each bin. Columns 3, 5 and 7 give unrectified luminosity functions, $LF$, whereas the luminosity functions corrected for incompleteness of photometry, $LFC$, are given in columns 4, 6 and 8. The subscripts $U$, $B$ and $V$ refer to the filter bandpasses.

| Min | Max | $LF_V$ | $LFC_V$ | $LF_B$ | $LFC_B$ | $LF_U$ | $LFC_U$ |
|------|------|------|------|------|------|------|------|
| 16.5 | 17.0 | 63 | 66 | 28 | 28 | 10 | 10 |
| 17.0 | 17.5 | 183 | 193 | 27 | 27 | 26 | 26 |
| 17.5 | 18.0 | 250 | 261 | 58 | 60 | 28 | 28 |
| 18.0 | 18.5 | 288 | 305 | 214 | 220 | 53 | 54 |
| 18.5 | 19.0 | 223 | 235 | 277 | 289 | 181 | 191 |
| 19.0 | 19.5 | 256 | 272 | 268 | 276 | 305 | 323 |
| 19.5 | 20.0 | 213 | 225 | 203 | 211 | 269 | 283 |
| 20.0 | 20.5 | 204 | 221 | 225 | 234 | 169 | 179 |
| 20.5 | 21.0 | 186 | 206 | 202 | 215 | 150 | 163 |
| 21.0 | 21.5 | 186 | 211 | 167 | 181 | 179 | 200 |
| 21.5 | 22.0 | 172 | 206 | 167 | 183 | 136 | 170 |
| 22.0 | 22.5 | 171 | 225 | 171 | 198 | 55 | 158 |
| 22.5 | 23.0 | 177 | 238 | 140 | 173 | - | - |
| 23.0 | 23.5 | 162 | 322 | 165 | 217 | - | - |
| 23.5 | 24.0 | 44 | 258 | 130 | 219 | - | - |
| 24.0 | 24.5 | - | - | 44 | 207 | - | - |

corresponds to $10.0 < M_V < 10.5$.

### 4.5. Binaries, blue/yellow stragglers and some comments on the cluster CMD

Color-magnitude diagrams of well studied open clusters frequently show "secondary sequences" located slightly above the "normal" main sequences, but offset by about 0.75 mag at a constant color (*e.g.* Anthony-Twarog et al. 1990; Bonifazi et al. 1990; Montgomery et al. 1993). There is a general agreement that the stars located above main sequence are binaries consisting of components with similar luminosities. It was shown by Bolte (1991) that large number of Praesepe stars located above main sequence of this cluster are spectroscopic binaries. A sequence of binary stars is present in the CMD of NGC 6791 shown in Fig. 10. This sequence is visible even better in Fig. 22 giving the field-star-corrected CMD for the central part of the cluster (field F1). This CMD was corrected statistically for the contaminating field population using the CMD for field F4 (see Section 2.1 and Fig. 10).

Besides stars located above the main sequence of NGC 6791, one may note the presence of several candidates for blue and yellow stragglers in Fig. 22. The total number of yellow/blue stragglers in NGC 6791 was estimated at about 100 by KU. Kaluzny and Rucinski (1993) discovered contact binaries among these stars. However, it is difficult to discuss in details properties of the NGC 6791 blue/yellow stragglers population without membership information. In fact, Rucinski (1994) pointed out that the alleged contact blue stragglers are probably foreground Milky Way interlopers belonging to a younger population than NGC 6791. Results of the deep proper-motion study being conducted by Cudworth (1994) are likely to shed a new light on this subject. Here, we would like to note only that the "cleaned" CMD of NGC 6791 contains several stars located a few tenths of magnitude above horizontal part of the subgiant branch. These stars are most probably binaries with mass ratios close to unity.

In addition to the blue/yellow stragglers, the "cleaned" CMD contains also a group of stars located below the main sequence of the cluster. The area in question is limited by $0.55 < B-V < 0.8$ and $18.5 < V < 20.0$. There are 42 and 21 stars located in this region on the CMDs for fields F1 and F4 respectively. The excess observed in field F1 in respect to field F4 is statistically significant at the $3\sigma$ level. We checked carefully that photometry for both fields is on the same photometric system and that observed effect is not due to incompleteness of photometry or presence of some spurious stars in the analyzed CMDs. The galactic latitudes of both fields differ by just 6′ and it is unlikely that we observe simply a gradient in the density of the field stars behind the cluster. Presence of several cluster members below the main sequence of the cluster CMD would be rather unexpected. It is tempting to relate these stars to the hot subdwarfs observed in NGC 6791. However, it is even more important to establish the membership status of these stars.

Table 7: Luminosity function for the "comparison" field located about 12′ North of center of NGC 6791 (field F4). First two columns define range of magnitudes for a given bin. Columns 3, 5 and 7 give unrectified luminosity function. The luminosity function corrected for incompleteness of photometry is given in columns 4, 6 and 8. The subscripts $U$, $B$ and $V$ refer to bandpasses.

| Min | Max | $LF_V$ | $LFC_V$ | $LF_B$ | $LFC_B$ | $LF_U$ | $LFC_U$ |
|------|------|------|------|------|------|------|------|
| 16.5 | 17.0 | 13 | 13 | 6 | 6 | 2 | 2 |
| 17.0 | 17.5 | 12 | 12 | 4 | 4 | 3 | 3 |
| 17.5 | 18.0 | 21 | 21 | 12 | 12 | 5 | 5 |
| 18.0 | 18.5 | 32 | 33 | 9 | 9 | 9 | 9 |
| 18.5 | 19.0 | 42 | 46 | 20 | 20 | 6 | 6 |
| 19.0 | 19.5 | 42 | 43 | 32 | 32 | 16 | 16 |
| 19.5 | 20.0 | 70 | 71 | 32 | 32 | 27 | 27 |
| 20.0 | 20.5 | 48 | 49 | 44 | 45 | 23 | 23 |
| 20.5 | 21.0 | 64 | 65 | 54 | 55 | 32 | 32 |
| 21.0 | 21.5 | 74 | 77 | 44 | 45 | 46 | 47 |
| 21.5 | 22.0 | 67 | 69 | 59 | 61 | 40 | 41 |
| 22.0 | 22.5 | 82 | 86 | 54 | 56 | 32 | 35 |
| 22.5 | 23.0 | 107 | 121 | 70 | 73 | 27 | 49 |
| 23.0 | 23.5 | 61 | 131 | 71 | 78 | 5 | 47 |
| 23.5 | 24.0 | 12 | 146 | 22 | 39 | - | - |
| 24.0 | 24.5 | - | - | 9 | 76 | - | - |

Let us return now to candidate binaries located above main sequence. In Fig. 23 we show a histogram of the color residuals relative to the main-sequence ridge (see Appendix B), for stars with $18 < V < 21$ from field F1. Figure 24 shows an analogous histogram for the field comparison field F4. The observed FWHM of the main sequence of NGC 6791 is about 0.05 mag. Taking into account observational errors it may be concluded, that the intrinsic width of the main sequence of the cluster does not exceed 0.025 mag. This allows to put a rather weak limit $\delta Z < 0.08$ on a range of metallicities exhibited by NGC 6791 stars.

There are 37 and 34 stars with $-0.155 < \delta(B-V) < -0.055$ in Figs. 23 and 24, respectively. Judging by this similarity in numbers, we conclude that most of the stars in this range of $\delta(B-V)$ belong to the field population. By assuming as single stars the objects with $-0.058 < \delta(B-V) < 0.058$ and as binaries those with $0.058 < \delta(B-V) < 0.200$, the data shown in Fig. 23 imply a lower limit to the fraction of binaries in the central part of NGC 6791 as 11 percent. This number is reduced to 10 percent, if we use data shown in Fig. 24 to make correction for the contaminating population of the field stars. This value is lower than percentage of "photometric" binaries observed in central regions of NGC 2420 (21%; Anthony-Twarog et al. 1990), NGC 2243 (30%; Bonifazi et al. 1990) and M67 (22%; Montgomery et al. 1993). Clearly, from the analysis of CMDs one can estimate only the relative frequency of main sequence binaries with mass ratios close to unity. However, the difference in relative frequency of binaries observed between NGC 6791 and three listed above clusters is significant. The likely cause of this difference is the relatively high mass of NGC 6791. KU estimated mass contained in stars with $V < 21.0$ ( $M_V < 7.5$ ) at about 4070 $M_\odot$ which is several times higher than estimated total masses of M67, NGC 2243 and NGC 2420. The observed frequency of binaries is modified during the lifetime of a given cluster by a process of mass striping. Stars with lower masses are more susceptible to this process (Henon 1969). Hence, binaries have a larger probability to survive as cluster members than single stars. Clusters with high mass are less susceptible to mass loss and, as a result, are likely to show lower (i.e. less modified) frequency of photometric binaries on their CMDs.

In Fig. 25 we show an expanded view of the "cleaned" CMD presented already in in Fig. 22. Some fine details of the upper main sequence and the subgiant branch of the cluster are visible. The intrinsic width of the main sequence increases as we move up from $V \approx 19$ to $V \approx 18$. There is also a hint for a narrow gap on upper main sequence at $V \approx 17.5$. The spread in the $B - V$ color observed for stars located at the base of the red giant branch ($17.5 < V < 16.4$) is larger than it may be expected from the errors of the photometry. In fact the lower section of the giant branch is wider than the upper main sequence of the cluster. This effect can hardly be explained by the contamination of the cluster CMD by the field stars (compared CMDs shown in Fig. 10). The most likely explanation of the observed spread in colors are differences in chemical abundances among NGC 6791 giants. Clearly, spectroscopic data are needed to resolve this problem.

## 5. Conclusions

We summarize the essential results of this paper:

1. We have obtained a substantial body of new photometry which can be used for comparison with stellar models.

2. The new photometry includes results for the $U$ filter and is thus particularly useful for detection of hot subdwarfs. Two new bright sdB/O candidates have been identified (stars B9 and B10).

3. We note the very narrow range of luminosities of the sdB/O stars: $V \approx 17.7 \pm 0.5$ and a striking defficiency of fainter sdB/O stars within $18 < V < 20$. An excess of faint blue stars with $V \approx 22$ is observed in the center of the cluster.

4. The blue star B8, which was identified first by KU, is likely an AM Her or U Gem-type cataclysmic variable.

5. The UBV photometry has been used to estimate reddening of the cluster. Comparison of observed $B - V$ colors of sdB/O stars belonging to NGC 6791 implies $E(B - V) = 0.17$.

6. Analysis of $U - B$ versus $B - V$ diagram indicates that metallicity of NGC 6791 is higher than solar. Lack of appropriate calibration for metal-rich stars hampers precise determination of metallicity of NGC 6791 through the use of the ultraviolet excess $\delta(U - B)$. Differential comparison of CMDs of NGC 6791 and M67 implies $[Fe/H] \approx +0.50$ for the former cluster.

7. NGC 6791 is by about 1 Gyr older than NGC 188. Detailed comparison of CMD of NGC 6791 with the new Padova isochrones (Bertelli et al. 1994) was hampered by unavailability of appropriate program, which would allow to interpolate between isochrones for different metallicities and ages.

8. The luminosity function of NGC 6791 is approximately flat for $6.5 < M_V < 10.5$.

9. We note a relatively weak secondary (binary) main sequence of NGC 6791 which implies a relatively lower frequency of equal-mass binaries relative to less populous old open clusters.

## *acknowledgments*

JK was supported by the Polish Committee of Scientific Research through grant 2P03D-008-08. SMR was supported by an operating grant from the Natural Sciences and Engineering Research Council of Canada.

## *Appendix A*

Tables containing photometry described in Sections 2.1 and 2.2 are published by Astronomy and Astrophysics at the Centre de Données de Strasbourg. See the Editorial in A&A 1993, Vol. 280, page E1.

Table 8: Fiducial points for NGC 6791. The entries for $B - V = 1.330$ and $B - V = 1.391$ describe location of the red giant branch clump.

| B-V | V | B-V | V | B-V | V |
|-----|-----|-----|-----|-----|-----|
| 1.330 | 14.58 | 1.164 | 17.42 | 0.912 | 18.35 |
| 1.391 | 14.59 | 1.123 | 17.52 | 0.933 | 18.50 |
| 1.638 | 13.77 | 1.104 | 17.50 | 0.962 | 18.70 |
| 1.579 | 14.03 | 1.043 | 17.47 | 1.019 | 19.09 |
| 1.420 | 14.69 | 0.981 | 17.41 | 1.068 | 19.40 |
| 1.349 | 15.33 | 0.948 | 17.39 | 1.095 | 19.60 |
| 1.341 | 15.42 | 0.920 | 17.40 | 1.145 | 19.80 |
| 1.312 | 15.69 | 0.896 | 17.50 | 1.205 | 20.10 |
| 1.273 | 16.02 | 0.874 | 17.59 | 1.277 | 20.50 |
| 1.243 | 16.37 | 0.877 | 17.78 | 1.326 | 20.70 |
| 1.213 | 16.75 | 0.887 | 17.96 | 1.388 | 21.00 |
| 1.192 | 17.18 | 0.891 | 18.10 | 1.450 | 21.27 |
| 1.183 | 17.34 | 0.901 | 18.20 | 1.483 | 21.50 |

Table 9: Fiducial points for NGC 188. The entries for $B - V = 1.190$ and $B - V = 1.210$ describe approximate location of the red giant branch. This table is based on photometry published by Kaluzny (1990) supplemented with unpublished data obtained at KPNO.

| B-V | V | B-V | V | B-V | V |
|-----|-----|-----|-----|-----|-----|
| 1.190 | 12.34 | 0.897 | 14.98 | 0.775 | 16.20 |
| 1.210 | 12.34 | 0.839 | 14.87 | 0.819 | 16.50 |
| 1.300 | 12.25 | 0.777 | 14.83 | 0.881 | 16.83 |
| 1.230 | 12.92 | 0.734 | 14.90 | 0.935 | 17.17 |
| 1.180 | 13.29 | 0.701 | 15.05 | 1.028 | 17.60 |
| 1.102 | 13.97 | 0.696 | 15.28 | 1.109 | 17.92 |
| 1.055 | 14.48 | 0.713 | 15.65 | 1.166 | 18.23 |
| 1.013 | 15.03 | 0.746 | 15.92 | 1.260 | 18.57 |
| 0.978 | 15.05 | 0.746 | 15.92 | | |

Figure 1: Residuals (derived - standard) for the standard stars observed on the night of October 6, 1991. Stars from the Landolt (1983) list are marked with squares.

Figure 2: The CMD for stars from the NGC 6791 region observed with the KPNO 2.1-m telescope + T1KA camera in October 1991. All stars detected in fields F1-4 are plotted.

Figure 3: Comparison of BV photometry of NGC 6791 based on 1991 KPNO data (T1KA chip) with photometry published by KU. Residuals are in the form (new - old) photometry.

Figure 4: A schematic chart for the field of NGC 6791 observed with the 0.9-m telescope at KPNO in October 1992. West is up and North is to the left. One pixels equals 0.68 arcsec. Each dot corresponds to a star with $V < 19$
.

Figure 5: V vs. B-V and V vs. U-B CMD's for stars from the NGC 6791 region based on photometry obtained with the KPNO 0.9-m telescope + T2KA camera in October 1992.

Figure 6: V vs. V-I CMD for stars from the NGC 6791 region based on photometry obtained with the KPNO 0.9-m telescope + T2KA camera in October 1992.

Figure 7: Comparison of BV photometry of NGC 6791 based on 1992 KPNO data with photometry published by KU. Residuals are in the form present minus old.

Figure 8: V vs. B-V and V vs. U-B CMD's for stars from the NGC 6791 region based on photometry obtained with the KPNO 2.1-m telescope + ST1K camera in October 1991. Stars with with $V < 18.5$ and photometry flagged as poor were not plotted. Average positions of variable stars B7 and B8 are marked with crosses.

Figure 9: Finding charts for the new blue stars B9 (left) and B10 (right). The faintest marked stars have $V \approx 20$. North is up and East is to the left. 100 pixels correspond to 68 arcsec.

Figure 10: The CMD for field F1 (cluster center) and field F4 (comparison field). This photometry is based on the data described in Sec. 2.1

Figure 11: The color-color diagram for turnoff stars with $17.5 < V < 18.9$ (circles), main sequence stars with $18.9 < V < 19.5$ (triangle) and giants (squares). The solid line represents relation adopted for Hyades.

Figure 12: The fiducial sequence for NGC 6791 (solid line) with the overplotted CMD for M67. Offsets of $\Delta V = 4.01$ and $\Delta(B-V) = 0.20$ were applied to the CMD of M67.

Figure 13: The fiducial sequence for NGC 6791 (solid line) with the overplotted CMD for M67. Offsets of $\Delta V = 3.84$ and $\Delta(B-V) = 0.12$ were applied to CMD of M67.

Figure 14: The fiducial sequence for NGC 6791 (solid line) with the overplotted CMD for M67. Offsets of $\Delta V = 3.88$ and $\Delta(B-V) = 0.05$ were applied to CMD of M67.

Figure 15: The fiducial sequences for NGC 6791 (solid line) and NGC 188 (dashed line). The filled squares denote positions of two helium burning giants in NGC 188. Offsets of $\Delta V = 2.21$ and $\Delta(B - V) = 0.178$ were applied to the NGC 188 sequence.

Figure 16: Lines of constant metallicity in the $log$(age) vs. $\Delta V$ plane. Positions of NGC 188 and NGC 6791 are marked with triangles

Figure 17: The observed V luminosity function for the central part of NGC 6791 (field F1; bottom) and for the nearby field F4 (top). $LF$ is the number of stars in intervals 0.5 mag wide.

Figure 18: Dependences of parameters $SIGMA$, $CHI1$ and $CHI2$, describing quality of photometry returned by DaOPHOT, on the V magnitude for the 600 sec exposure of field F4. Objects with $CHI1 > 2$ or $1 < CHI2$ or $CHI2 < -1$ were excluded while deriving luminosity function. Also objects with atypically large values of $SIGMA$ for their magnitude (triangles) were excluded from analysys. Only objects with $CHI1 < 2$ and $-1 < CHI2 < 1$ are shown on the $V$ vs. $SIGMA$ plot.

Figure 19: The rectified luminosity functions in the UBV bands for the central part of NGC 6791 (field F1). $LF_c$ is the number of stars in intervals 0.5 mag wide.

Figure 20: The rectified luminosity functions in the UBV bands for the field F4 located about 12´ from the center of NGC 6791. $LF_c$ is the number of stars in intervals 0.5 mag wide.

Figure 21: The rectified and field-stars-corrected luminosity function for the central part of NGC 6791. $LFc$ is the number of stars in intervals 0.5 mag wide.

Figure 22: The field-stars-corrected CMD for the central part of NGC 6791. Large fraction of stars located slightly above cluster main sequence are most probably binaries.

Figure 23: The histogram of the $B - V$ residuals relative to the main sequence ridge line, for stars with $18 < V < 21$ from the field F1.

Figure 24: The histogram of the $B - V$ residuals relative to the main sequence ridge line, for stars with $18 < V < 21$ from the field F4. The scales are the same as for Fig. 23.

Figure 25: The expanded view of the field-star corrected CMD (see Fig. 22) near the turnoff.